# Non-resonant lasing in a deep-hole scattering cavity

CHULMIN OH,[1,2,†] HO JIN MA,[3,4,†] KYEOREH LEE,[1,2,5] DO KYUNG KIM,[4,6] AND YONGKEUN PARK[1,2,7]

[1]*Department of Physics, Korea Advanced Institute of Science and Technology (KAIST), Daejeon, 34141, Republic of Korea*
[2]*KAIST Institute for Health Science and Technology, KAIST, Daejeon, 34141, Republic of Korea*
[3]*Department of Engineering Ceramics, Korea Institute of Materials Science, Changwon, 51508, Republic of Korea*
[4]*Department of Materials Science and Engineering, KAIST, Daejeon, 34141, Republic of Korea*
[5]*kyeo@kaist.ac.kr*
[6]*dkkim@kaist.ac.kr*
[7]*yk.park@kaist.ac.kr*

**Abstract:** Random lasers are promising in the spectral regime, wherein conventional lasers are unavailable, with advantages of low fabrication costs and applicability of diverse gain materials. However, their practical application is hindered by high threshold powers, low power efficiency, and difficulties in light collection. Here, we demonstrate a power-efficient easy-to-fabricate non-resonant laser using a deep hole on a porous gain material. The laser action in this counterintuitive cavity was enabled by nonresonant feedback from strong diffuse reflections on the inner surface. Additionally, significant enhancements in slope efficiency, threshold power, and directionality were obtained from cavities fabricated on a porous Nd:YAG ceramic.

## 1. Introduction

A random laser is a peculiar type of laser that utilizes multiple scattering to trap light [1-3]. The narrow-spectrum emission from a disordered gain medium, particularly when related to mesoscopic interference effects in complex scattering media [4], has intrigued researchers. Furthermore, the practical advantages of random lasers have been explored. Random lasers do not require transparent gain media with a carefully aligned laser cavity; diverse gain media, such as ceramics [5], powders [6-8], and films [9], have been introduced, which provide variegated available lasing frequencies, including ultraviolet [8] and terahertz [10, 11].

Despite the advantages of laser gain media, random lasers have seldom been utilized in practical applications. The first reason is its inherently low power efficiency. In most random lasers, multiple scattering not only traps the emission light but also rejects pumping light effectively [12]. Although one- [13] and two-dimensional [11, 14, 15] random lasers can avoid this issue by separating the pumping and lasing dimensions, confining the direction of multiple scattering to lower dimensions is not always possible. Second, random lasers have low directionality (or low spatial coherence). Although temporally coherent but spatially incoherent light is advantageous for specific applications, such as speckle-free coherent imaging [16], it is typically challenging for random lasers to replace conventional lasers in most applications. Random lasers with controllable directionality have been studied to address this issue [11, 14, 17].

Recently, a scattering cavity was introduced within the scattering gain medium to improve the power efficiency and directionality of random lasers [18]. Inspired by a fish trap, the cavity is composed of a spacious internal volume and small entrance [Fig. 1(a)]. Owing to the easy-to-get-in but hard-to-get-out structure, the pumping light is eventually absorbed into the gain medium. Contrastingly, the emitted light is successively amplified through reflections from the cavity wall made of a gain medium. When the gain exceeds the scattering loss, it becomes a laser. Unlike conventional lasers, this laser cannot retain a stable resonant mode in the cavity owing to random scattering and is called a non-resonant laser (NRL) to be distinguished from random lasers without an explicit cavity [19, 20]. Notably, most random lasers are NRLs, except random lasers with coherent feedback [21].

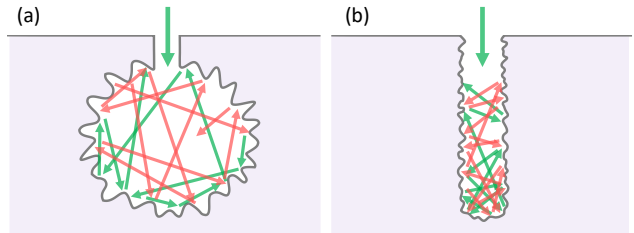

Fig. 1. Two cavity designs for non-resonant lasers. Light trapping is enhanced by diffuse reflection on the cavity wall in both cavities: (a) a spherical vacancy with a small outlet, (b) a vertical hole with a sufficient depth. The green and red lines indicate the pump and emission light, respectively.

The incorporation of a scattering cavity successfully enhances the power efficiency and directionality of the laser. However, carving a spherical cavity within a scattering medium requires a complicated fabrication process. Because a simple configuration is one of the major advantages of random lasers, such practical difficulties severely undermine their merits and hinder their popularization. Furthermore, the lasing properties related to the cavity scale, such as the lasing threshold and spatial coherency, are practically limited as the fabrication difficulty increases for smaller cavities.

To mitigate these practical difficulties, we introduced a deep hole on the surface of the scattering medium as a laser cavity [Fig. 1(b)]. Even with an open-top cavity structure, efficient light trapping can be realized based on diffuse reflection that randomly redirects light. Deep-hole cavities were fabricated by drilling a micro-sized hole on the surface of a porous Nd:$Y_3Al_5O_{12}$ (Nd:YAG) ceramic. The successful operation of NRLs in deep-hole cavities was enabled by the introduction of pumping light within the cavities. Significant improvements in directionality, threshold, and power efficiency were observed compared to those in a previous work [18]. The lasing characteristics at various cavity diameters and depths were explored and discussed in comparison with numerical results.

## 2. Principle

Unlike a spherical cavity, a deep-hole cavity does not have a narrower entrance than its interior space. The straight and open-top structure appears ineffective for light trapping, and this intuition is true if the cavity wall is a mirror surface. However, we found that the exit probability diminishes dramatically when the cavity is made of a scattering medium with a diffusely reflecting surface. Assuming that the cavity wall is a perfect diffuser, the direction of the reflected light obeys Lambert's cosine law independent of its incident direction [22] [Fig. 2(a)]. Because every wall reflection randomly redirects the light trajectory, the internal-cavity photons can be regarded as one-dimensional random walkers along the depth direction, which has an rapidly decreasing probability distribution with increasing displacement. The random walk scheme suggests a rapidly increasing trapping ability with an increase in the aspect ratio (AR) of the cavity.

To visualize the trapping ability of the deep-hole cavities, we numerically calculated the photon trajectories inside the cavities [Fig. 2(b)]. The gain or loss from the wall reflections has been excluded here for simplicity, without losing generality. One hundred photons were introduced at the bottom of cavities with various ARs (1, 3, and 5), and the better trapping ability for larger AR was well presented by the denser photon trajectories. A higher photon density for a deeper cavity depth was observed, which was also anticipated from the random walker analogy.

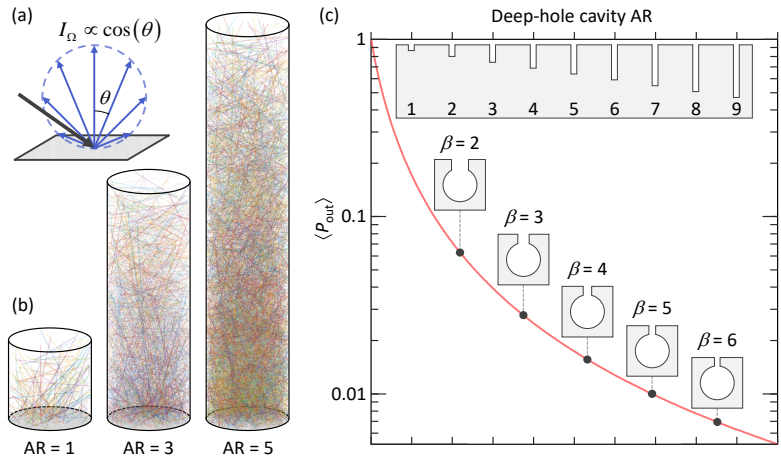

Fig. 2. Light trapping capability of the proposed cavity. (a) Illustration of Lambert's cosine law. The distribution of radiant intensity after reflection ($I_\Omega$) is proportional to the angle of reflection ($\theta$), independent of the incident angle. (b) Overlapped trajectories of photons in the deep-hole cavities with different aspect ratios (ARs). Light trajectories at each AR are accumulated for hundred photons departing from the cavity bottom, and diffusely reflecting until they exit the cavity. Higher AR yields the denser and longer trajectories. (c) The expected probability that a photon directly leaves a deep-hole cavity after a reflection (red line). The probability drops drastically as AR increases. Spherical cavities and corresponding exit probability (black dots) are plotted together. $\beta$ indicates the diameter ratio of the cavity to the entrance. The $y$-axis is in log scale.

To quantify the trapping ability of the deep-hole cavity, we calculated the expected exit probability for each reflection, $\langle P_{out} \rangle$ [Fig. 2(c)]. We found that $\langle P_{out} \rangle$ diminished dramatically as AR increased. This is a consequence of the increasing photon density and decreasing exit probability with an increase in the cavity depth. The cavity with AR = 2 already provides $\langle P_{out} \rangle < 0.1$, implying that a photon experiences >10 wall reflections on average before it leaves the cavity. To compare the trapping abilities directly, spherical cavities that provide the same $\langle P_{out} \rangle$ have been depicted over the graph with the corresponding cavity-to-entrance diameter ratio ($\beta$). Because successful NRLs were demonstrated with spherical cavities of $\beta = 2$–4 [18], Fig. 2(c) may provide a rough reference for suitable AR ranges in NRLs. However, the above numerical results do not consider the gain and loss from wall reflections, which are important parameters for the lasing threshold and efficiency. To properly estimate the lasing performance, a more complex numerical model that considers uneven distributions of pumping light, emission light, and gain within the cavity is required (see Supplement 1).

## 3. Materials and methods

In the experiments, we fabricated deep-hole cavities by micro-drilling porous Nd:YAG bulk ceramics. To explore lasing characteristics, deep-hole cavities with various diameters and ARs were engraved on a single ceramic. A coupling lens was used to deliver pumped light to the bottom of the cavities and collimate laser emission.

### 3.1 Synthesis and characterization of Nd:YAG powders

To obtain high-purity Nd:YAG particles, a co-precipitation method was used. The starting precursors were yttrium nitrate tetrahydrate ($Y(NO_3)_3 \cdot 4H_2O$, $\geq$ 99.99%, Sigma-Aldrich Inc., USA), neodymium nitrate hexahydrate ($Nd(NO_3)_3 \cdot 6H_2O$, $\geq$99.9%, Sigma-Aldrich Inc., USA), and aluminum ammonium sulfate dodecahydrate ($NH_4Al(SO_4)_2 \cdot 12H_2O$, reagent grade, Alfa Aesar Inc., USA). The raw materials were homogeneously mixed and dissolved in deionized (DI) water at 40°C. In this study, the concentration of $Nd^{3+}$ ions incorporated into the YAG phase was fixed at 1 at.%. For precipitation, a solution was prepared by dissolving ammonium bicarbonate ($NH_4HCO_3$, $\geq$ 99%, Sigma-Aldrich Inc., USA) in a mixed solvent of ethanol and DI water. The volume ratio of ethanol to DI water was 0.6. The precursor solution was added dropwise to the precipitant solution at a dripping rate of 3 ml/min at room temperature. The suspension was aged for 24 h, centrifuged, and washed repeatedly with water and ethyl alcohol in sequence to obtain the precipitate. It was then dried at 85°C for 24 h in an oven. To achieve crystallization and eliminate residual organics, the particles were calcined in air at 1,250°C for 4 h. The calcined powders were sieved through a 200-mesh screen.

### 3.2 Fabrication of Nd:YAG porous bulk ceramic

The synthesized powders were pelletized by uniaxial pressing at 30 MPa into a 6-mm-diameter stainless steel mold with a sample thickness of 2 mm. The green bodies were then cold isostatic pressed (CIP) at 200 MPa for 5 min. To prepare porous Nd:YAG pellets, they were sintered at 1,350°C for 10 h at a heating/cooling rate of 5°C/min in air. The consolidated Nd:YAG ceramics were characterized using conventional X-ray diffraction (XRD) with Cu Kα radiation at a scan rate of 5°/min between 10° and 80° (SmartLab, Rigaku Inc., Japan). The observed X-ray diffraction patterns were consistent with the standard diffraction peaks for the cubic YAG phase (JCPDS #33-0040) [Fig. 3(a)]. Microstructure images of the densified Nd:YAG ceramics were obtained by scanning electron microscopy (SEM, Model Philips XL 30 FEG, Koninklijke Philips N.V., Netherlands) after diamond polishing [Fig. 3(a), inset]. The average domain size of the consolidated sample was determined by multiplying the linear-intercept length of 200 grains by 1.56 [23]. Bulk density was measured using Archimedes' method. Samples with a relative density of 70% and an average domain size of 400 nm were obtained because the sintering temperature was much lower than that of conventional Nd:YAG laser ceramics with full density.

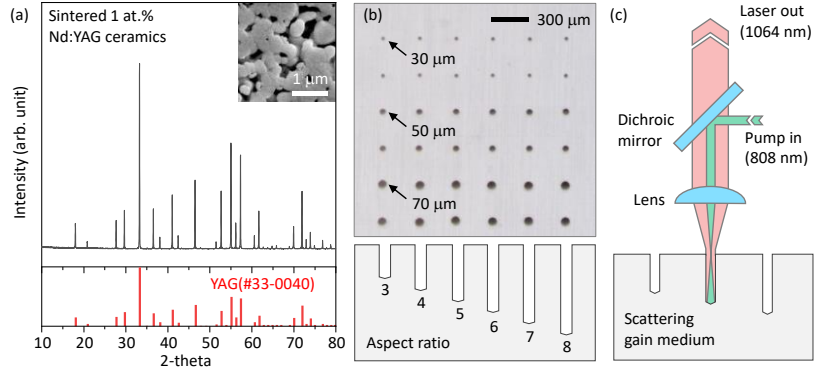

Fig. 3. The proposed laser system (a) X-ray diffraction (XRD) result of the sintered Nd:YAG ceramics (black) and standard diffraction peaks of cubic YAG phase (red), showing fair agreement between measured results and expected peaks. The inset is a scanning electron microscopy image of the microstructure of the ceramic. (b) Arrangement of cavities fabricated on the Nd:YAG ceramic pellet. ARs increase from left to right (3, 4, 5, 6, 7, and 8) while diameters increase from top to bottom every two rows (30, 50, and 70 μm). (c) The optical system. A coupling lens focuses pump light into the bottom of cavity fabricated on the ceramic pellet and collimates the emission light. A dichroic mirror is placed to separate the pumping and emission light.

### 3.3 Fabrication of deep-hole cavities

The surfaces of the fabricated Nd:YAG ceramic pellets were polished and attached to a Cu holder for machining. Deep-hole cavities were fabricated using a commercial microdrilling machine (G. I. Tech Co., Ltd, Cheonan, Republic of Korea). Two copies of cavities of three diameters (30, 50, and 70 μm) with six different ARs (3, 4, 5, 6, 7, and 8) –total 36– were engraved on the same surface of the ceramic pellet [Fig. 3(b)]. The bottom faces of the drilled cavities were indented at 118°, owing to the point angle of the drill bits. After machining, the ceramic holder assembly was immersed in acetone and sonicated for 30 min to detach the ceramic from the holder and remove debris. Subsequently, the ceramic was baked again at 800 °C for 2 h to remove potential organic contamination during the cavity fabrication process.

### 3.4 Optical setup

The experimental setup comprises a coupling lens, dichroic mirror, and pump source, which is only necessary for the laser action of the proposed cavity [Fig. 3(c)]. The coupling lens ($f$ = 20.0 mm, NA = 0.52, AL2520H-B, Thorlabs, Inc.) was used to deliver pumping light into the cavity and to collimate the emission light. A dichroic mirror (FF875-Di01-25 × 36, Semrock, Inc.) was used to separate the laser emission from pumping light. A femtosecond Ti: sapphire laser (> 2 W, FWHM = 5 nm at 808 nm, Chameleon Vision-S, Coherent Inc.) was used as the pumping source. The delivered pump power was calibrated by measuring the optical power transmitted through a dichroic mirror.

We measured the power and spectrum of laser emission as a function of the delivered pump power. A silicon photodiode (S121C, Thorlabs, Inc.) with a bandpass filter ($\lambda_c$ = 1064 nm, $\Delta\lambda_{FWHM}$ = 10 nm, FL1064-10, Thorlabs, Inc.) was used for power measurements. An optical spectrum analyzer ($\delta\nu$ = 7.5 GHz, $\delta\lambda$ = 28.4 pm, at 1064 nm, OSA201C, Thorlabs, Inc.) was used for the spectrum measurements.

## 4. Results and discussion

### 4.1 Directionality (spatial coherency)

Because NRLs do not have resonant modes that constrain the spatial lasing profile, the laser emission is as spatially incoherent as possible within the constraints of a given beam diameter and divergence. Thus, the directionality of NRLs can be simply quantified by the number of available (incoherent) spatial modes, $N_x = \left(\frac{\pi}{2} \frac{NA}{\lambda} D\right)^2$, where $D$ is the outlet diameter, and NA is the numerical aperture of the coupling lens [24]. A smaller $N_x$ signifies a more directional laser emission after the coupling lens. For a given coupling lens (NA = 0.52), $N_x$ = 530, 1473, and 2888 for $D$ = 30 μ, 50 μ, and 70 μm, respectively. These values are much smaller than the $N_x$ values of the spherical cavities (21800 in Ref. [18]) and random lasers (~$10^7$ in Ref. [5]). Notably, the directionality of the NRL is related to the laser output power owing to the NA dependence of $N_x$. For example, if we decrease the NA of the coupling lens, the directionality of the laser emission increases (as $N_x$ decreases), whereas the laser output power decreases.

## 4.2 Spectral properties

The laser action was confirmed by narrowing the emission spectrum [Fig. 4(a)]. The uneven stimulated emission cross-section caused the narrowing of the linewidth during successive amplifications in the cavity, leaving only a single peak in the emission spectrum [25]. Accordingly, we observed the narrowing of the strongest peak of the Nd:YAG photoluminescence spectrum (1064 nm) as the pump power increased, while the other peaks (e.g., 1061.5 nm) were suppressed. For instance, the linewidth narrowed to 53 pm for a 50-μm-diameter cavity with AR = 8 at 1.0 W pump power.

Narrower linewidths were observed for cavities with larger ARs [Fig. 4(b)] and diameters [Fig. 4(c)]. However, if the pump power increased further, the linewidth was again broadened. Such linewidth rebroadening was observed regardless of cavity geometry, but a lower pump power was required to reach the pivot point for a cavity with a smaller AR or diameter.

Consistently increasing peak wavelengths were observed as the pump power increased [Figs. 4(d) and 4(e)]. We expect that the thermal effect is responsible for such a redshift in the spectrum [26]. A more significant redshift was observed for cavities with smaller ARs or diameters, which is reasonable as the pump power is localized to smaller volumes. For instance, the estimated temperature of the cavity from the redshifted peak was 305°C for the 30-μm-diameter cavity with AR = 3 at 1.6 W pump power. We suspect that this temperature rise may also be responsible for the linewidth rebroadening.

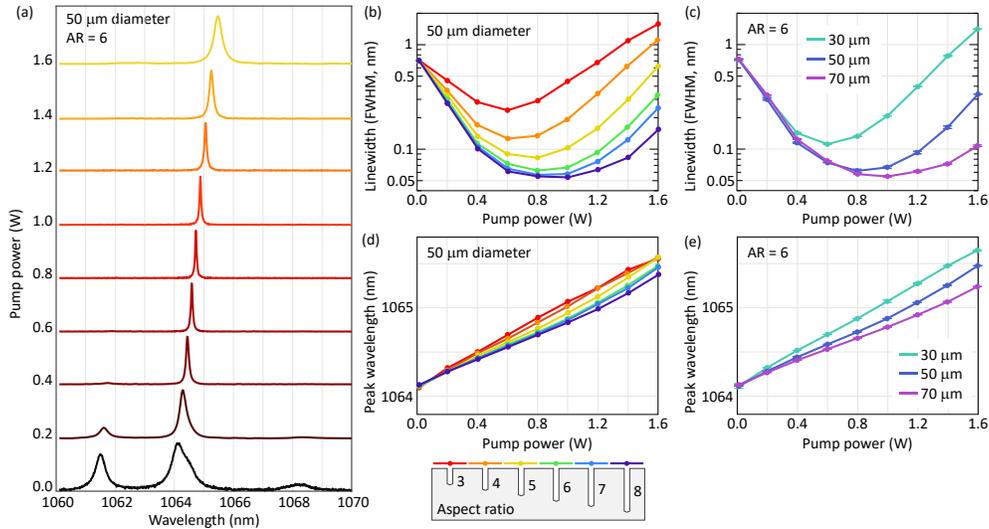

Fig. 4. Spectral properties for cavities with different aspect ratios (ARs). (a) Normalized spectra for a 50-μm-diameter cavity with AR = 6. The corresponding pump powers are denoted on the left. (b) Linewidths and (d) peak wavelengths of 50-μm-diameter cavities with different ARs. (c) Linewidths and (e) peak wavelengths of AR = 6 cavities with different diameters. Each error bar in (c, e) represents the minimum and maximum measured values obtained from two identical cavities.

## 4.3 Power properties

The laser output powers for various deep-hole cavities are shown in Fig. 5. The output power curves exhibit a smooth transition near the lasing threshold, as in random lasers [27, 28]. This is due to the spatial and spectral overlap between spontaneous and stimulated emissions [29]. After the threshold, the output power increased linearly, as in conventional lasers. However, as the pump power increased further, we observed that the slope efficiently decreased and eventually became negative. Such an output power drop generally occurs at lower pump powers, as the cavity diameter or ARs decrease. Based on the similarity to the aforementioned linewidth rebroadening, we also suspect that the temperature rise issue is one of the reasons for the output power drop.

Although the aforementioned trends were similar, the measured output powers for a given pump power were highly dependent on the cavity diameters and ARs (Fig. 5). For instance, we observed monotonically increasing slopes for increasing ARs in 30-μm-diameter cavities, while the slopes reached a maximum around AR = 5–6 in 50- and 70-μm-diameter cavities.

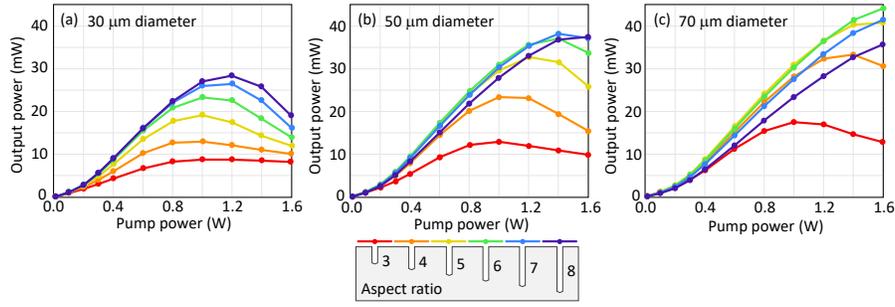

Fig. 5. Output powers for (a) 30-μm-, (b) 50-μm-, and (c) 70-μm-diameter cavities with different aspect ratios.

To compare the lasing power efficiencies, the slope efficiencies were quantified from the linearly increasing region [Fig. 6(a)]. Moreover, we estimated lasing thresholds by extrapolating the linear region. This is marked by triangles on the bottom line of Fig. 6(a). We achieved slope efficiencies of 3.40%, 3.95%, and 3.76% and lasing thresholds of 143 mW, 159 mW, and 174 mW for the 30-, 50-, and 70-μm-diameter cavities with AR = 6, respectively.

From the observations in Fig. 6(a), the deep-hole cavity design shows much lower threshold powers and even higher slope efficiencies than the previous spherical cavity design [18]. A reduced laser threshold is expected as the pumping rate per unit gain volume is increased by reducing the cavity size. The improved slope efficiency is an unexpected advantage derived from the open-top design, which allows direct free-space coupling, whereas the spherical cavity requires an optical fiber that introduces additional coupling losses.

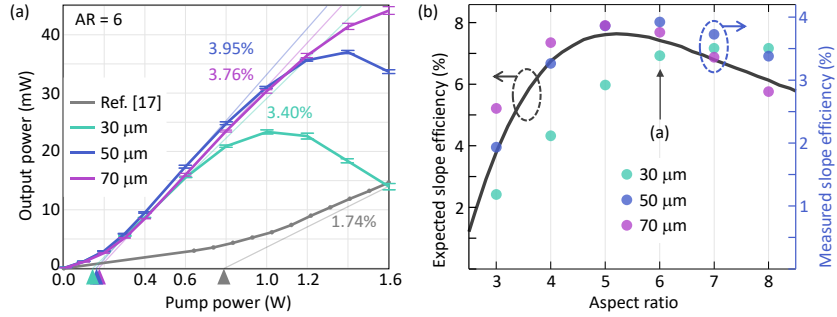

Fig. 6. (a) Output powers for cavities of aspect ratio (AR) = 6 with different diameters with the best result of spherical cavities in Ref. [18] (gray line). The linear regions of the power curves are extrapolated (dashed line) with their *x*-intercepts marked as triangles. The slope efficiencies estimated from the slope of each extrapolated line are shown near the line. The lasing thresholds estimated from the *x*-intercepts are 143 (cyan, 30 μm diameter), 159 (blue, 50 μm diameter), 174 (magenta, 70 μm diameter), and 790 mW (black, Ref. [18]), respectively. Each error bar represents the minimum and maximum measured values obtained from two identical cavities. (b) Slope efficiencies obtained from the numerical model (black line) and the experiment (dots). The left *y*-axis is for the slope efficiencies predicted by the numerical model while the right *y*-axis is for those measured in the experiment.

The quantified slope efficiencies for the various deep-hole cavities are shown in Fig. 6(b). Furthermore, we provide an expected slope efficiency curve calculated from the numerical model described in Supplement 1. The slope efficiencies predicted by the numerical model were independent of the diameter within the 30–70 μm range. The maximum slope efficiency around AR = 5–6 is expected from the numerical model, which is consistent with the experimental results for the 50- and 70-μm-diameter cavities. We suspect that the inconsistency with the results from the 30-μm-diameter cavities, and lower values than expected could be derived from various sources, such as temperature-dependent material properties and potential contamination during the fabrication steps.

## 5. Conclusion

In this study, we propose and demonstrate an NRL using a deep-hole cavity. We show that the deep-hole geometry outperforms the spherical geometry because it is not only considerably easier to fabricate but also provides a lower lasing threshold and better slope efficiency. This scalable cavity design allows controlling the lasing threshold, laser linewidth, and number of spatial modes (i.e., directionality) by adjusting the cavity diameter and depth. In this sense, our laser is promising for speckle-free imaging [16, 30], holographic display [31, 32] and quantitative phase imaging [33, 34], in which the spatiotemporal coherence of light sources is carefully controlled to balance the resolution and speckle reduction [35]. Nonetheless, an adequate temperature control scheme is required to provide better and more stable lasing. Recently, a wide variety of new materials, including nanoparticles [36], graphene [37], and perovskites [38-40], have been explored for random lasers. We expect the deep-hole geometry to be easily applicable to these materials by either fabricating a hole on the surface or depositing the material on a hole structure, which significantly enhances lasing characteristics.

**Funding.** This work was supported by the KAIST UP program, BK21+ program, Tomocube Inc., National Research Foundation of Korea (2015R1A3A2066550, 2021R1C1C2009220), KAIST Institute of Technology Value Creation, Industry Liaison Center (G-CORE Project) grant funded by the Ministry of Science and ICT (N11210014, N11220131), and Institute of Information & Communications Technology Planning & Evaluation (IITP; 2021-0-00745) grant funded by the Korean government (MSIT).

**Acknowledgments.** We thank Geumil Jang (G.I Tech Co., Ltd, Cheonan, Republic of Korea) for machining holes on the ceramics based on our design.

**Disclosures.** The authors declare no conflicts of interest.

**Data availability.** Data underlying the results presented in this paper are not publicly available at this time but may be obtained from the authors upon reasonable request.